\documentclass[intlimits,twoside,a4paper]{article}

\usepackage[cp1251]{inputenc}

\usepackage[eqsecnum]{cmpj3}

\usepackage{xcolor}

\newcommand{\bc}{\begin{center}}
\newcommand{\ec}{\end{center}}
\newcommand{\be}{\begin{equation}}
\newcommand{\ee}{\end{equation}}
\newcommand{\bea}{\begin{eqnarray}}
\newcommand{\eea}{\end{eqnarray}}

\newcommand{\RNum}[1]{\lowercase\expandafter{\romannumeral #1\relax}}

\issue{2023}{26}{4}{43703}
\doinumber{10.5488/CMP.26.43703}
\title[The role of intrinsic atomic defects in a Janus MoSSe/XN (X = Al, Ga) heterostructure: a first principles study]
{The role of intrinsic atomic defects in a Janus MoSSe/XN (X = Al, Ga) heterostructure: a first principles study}
  
\author[\"{O}. C. Yelgel]{\"{O}. C. Yelgel\orcid{0000-0001-5888-5743}}
  
\address{Recep Tayyip Erdo\~{g}an  University, Department  of  Electrical and Electronics Engineering, 53100, Rize, Turkey}


\date{Received August 4, 2023, in final form September 22, 2023}
\begin{document}
	
	\maketitle
	
\begin{abstract}
The interactions between different layers in van der Waals heterostructures have a significant impact on the electronic and optical characteristics. By utilizing the intrinsic dipole moment of Janus transition metal dichalcogenides (TMDs), it is possible to tune these interlayer interactions. We systematically investigate structural and electronic properties of Janus MoSSe monolayer/graphene-like Aluminum Nitrides (MoSSe/g-AlN) heterostructures with point defects by employing density functional theory calculations with the inclusion of the nonlocal van der Waals correction. The findings indicate that the examined heterostructures are energetically and thermodynamically stable, and their electronic structures can be readily modified by creating a heterostructure with the defects in g-AlN monolayer. This heterostructure exhibits an indirect semiconductor with the band gap of 1.627 eV which is in the visible infrared region. It can be of interest for photovoltaic applications. When a single N atom or Al atom is removed from a monolayer of g-AlN in the heterostructure, creating vacancy defects,  the material exhibits similar electronic band structures with localized states within the band gap which can be used for deliberately tailoring the electronic properties of the MoSSe/g-AlN heterostructure. These tunable results  can offer exciting opportunities for designing  nanoelectronics devices based on MoSSe/g-AlN heterojunctions.
 \printkeywords
\end{abstract}

\section{Introduction}

The rapid growth of technology leads to a rise in scientific studies on the micro-miniaturization and functioning of electronic devices. There are two key crucial aspects for the material to be utilized or discovered for this reason: first, it needs to have the necessary band gap or be able to modify its band gap; and second, it must be in nanocrystal form with carrier mobility. Because of its extraordinary physical and chemical characteristics, the discovery of single-layer graphene from graphite via the mechanical exfoliation method in 2004 \cite{1} has piqued the interest of several investigations on two-dimensional (2D) materials. Due to its unique electrical structure, massless carrier mobility, and rapid charge migration rates, graphene performs outstandingly \cite{2}. However, the lack of intrinsic band gap of graphene limits its direct use in nanoscale devices such as photocatalysis and field effect transistors (FETs)~\cite{3}. As a result, much emphasis has been placed on narrowing the band gap of graphene while retaining its interesting physical features. There is a strong incentive to examine and explore 2D graphene-like materials for the continued development of nanoscale devices due to their intriguing physical and chemical features \cite{4,5} that cannot be observed in bulk materials (for example, monolayer MoS2 possesses a direct band gap of 1.8 eV \cite{6}). Due to its  compositional flexibility and tunability, layered transition metal dichalcogenides (TMDs) have unique electrical, thermal, thermoelectric, optical, and catalytic characteristics \cite{7,8,9,10,11,12,13,14,15}. Numerous diverse TMDs structures can be designed in the form  of MX$_2$ (M = Mo, W, Nb, and/or V; X=S, Se, and/or Te)  with  impressive  transport   properties  by  tunable  band  gaps  \cite{16,17,amin}.   Thereby, TMDs  open  a new  horizon for  technological research developments in many  fields of FETs, rechargeable batteries, and electrochemical  and photocatalytic  H production from  H$_2$O.  Very recently, the  new derivatives of  TMDs with Janus structure  with the chemical formula  of MXY (M = Mo,  W; X, Y=S,  Se, Te), in  which Janus TMDs monolayer  sandwiched by two different  chalcogen layers, attract tremendous research  attention and offer new  opportunities for future applications  due to  their unique  electronic properties,  stability, optical and catalytic features sourced  from their mirror asymmetry in structure \cite{18,19}.   Different kinds of  methods can be  used for experimetal synthesis of 2D materials  such as atomic layer deposition (ALD), molecular beam epitaxy  (MBE), chemical vapor deposition (CVD), liquid  exfoliation   (LE),  physical  vapour  deposition   (PVD)  and micromechanical exfoliation (ME) \cite{20,21,22,23,24,25}.  The unique Janus  family  do  not  exist   in  nature  and  can  be  successfully synthesized using  the CVD  method~\cite{18,27}.   Previous studies proved that  Janus monolayers can be  promising application candidates in   the  fields   of  photocatalysts   \cite{28,29},  optoelectronics \cite{30},  thermoelectrics \cite{31},  valleytronics \cite{32,33,34}, gas sensing \cite{35}, and  spintronics \cite{36}.  Motivated by these promising and  unique properties of  Janus family, the  fabrication of heterostructured  Janus   materials  such  as   MoSSe/WSSe  \cite{37}, MoSSe/WSe$_2$~\cite{38}, and MoSSe/graphene \cite{39} was presented in the literature to expand the applications of these materials to more diverse fields.

MoSSe/AlN heterostructures have emerged as promising materials for various practical applications, because they were found to be energetically, dynamically, and thermally stable. Extensive theoretical studies were conducted on defect-free MoSSe/AlN heterostructures, revealing their indirect band gaps ranging from 1.00 eV to 1.68 eV \cite{prog}. Despite these band gaps being smaller than the desired threshold of 1.23 eV, the band edge positions demonstrate a favorable alignment for water splitting processes. This intriguing feature indicates that MoSSe/AlN heterostructures exhibit excellent activity in the visible-infrared region, making them suitable for efficient energy conversion and utilization \cite{prog}. An important aspect contributing to the superior performance of MoSSe/AlN heterostructures is the charge transfer between the MoSSe and AlN layers, which induces a strong built-in electric field. This electric field further enhances the separation of photogenerated carriers, thereby significantly improving the overall performance of the heterostructures \cite{jod}.

Fascinated by these intriguing discoveries in 2D heterostructured materials, we used first-principles studies to investigate the effects of intrinsic atomic defects on the structural and electrical characteristics of a MoSSe/g-AlN van der Waals heterostructure. These defects are produced by removing a single Al and/or N atom close to the interfaces. We first assess the stacking types and possibilities for producing MoSSe/g-AlN heterostructures by taking into account both the MoS and MoSe sides interacting with the monolayer g-AlN. We want to uncover the influence of intrinsic atomic defects on the electronic characteristics of MoSSe/g-AlN heterostructures after determining the most stable stacking type of MoSSe/g-AlN heterostructures. Then, based on the type of vacancy, more emphasis is laid on localized band gap states. The results show that the MoSSe/g-AlN heterostructure has many novel properties, such as a moderate band gap, appropriate band edge positions, and electronic-hole separation, which provide additional insights into the structural and electronic properties of MoSSe/g-AlN heterojunctions for use in nano/optoelectronic and spintronic applications.

\section{Computational method}

In this  work, all density  functional theory (DFT)  calculations were performed  using Quantum  Espresso  package \cite{espresso}  based  on DFT within plane-wave basis.  To describe  electron-ion interaction we  used norm conserving  pseudo-potentials \cite{normc},  which are  more accurate  than the ultra-soft pseudo-potentials. The  description of exchange-correlation interaction   was  performed using  the   Perdew-Burke-Ernzerhof  (PBE) functional  within   the  generalized  gradient   approximation  (GGA) \cite{pbe}. The contribution from the van der Waals (vdW) interaction between layers  was  included  using  the  DFT-D3  method  \cite{d3}.  The  fully relativistic  norm-conserving   pseudo-potentials  were   employed  to include spin-orbit coupling (SOC) in self-consistent calculations. The kinetic energy  cutoff  for  the plane  wave expansion  was set  to  2721 eV. For the  MoSSe/g-AlN heterostructure with defects, we used  4$\times$4 supercell. For the defect calculations we employed 80-atom supercells since it was found that they give a good convergence with regard to the system size.  The  integrations  over   the  Brillouin  zone  (BZ)  were performed  by  using  18$\times$8$\times$1 and 3$\times$3$\times$1   Monkhorst-Pack  k-point  mesh for perfect supercell and  the supercell with the defects, respectively \cite{monk}. To  hinder the interactions  in the  stacking direction between the repeated  slabs,  a  vacuum  layer  thickness  of  25  {\AA}  which gave well-converged results was added. By adding this vacuum region, the unit cell length of $c$ becomes 30~{\AA}. All   the  considered   structures  were  relaxed   until  the Hellmann-Feynman forces  acting on every  atom were smaller  than 0.01 eV/A while the total energy convergence was set to below 10$^{-6}$ eV.

\section{Results and discussion}
\begin{figure}[!t]
\centering{\includegraphics[width=12cm]{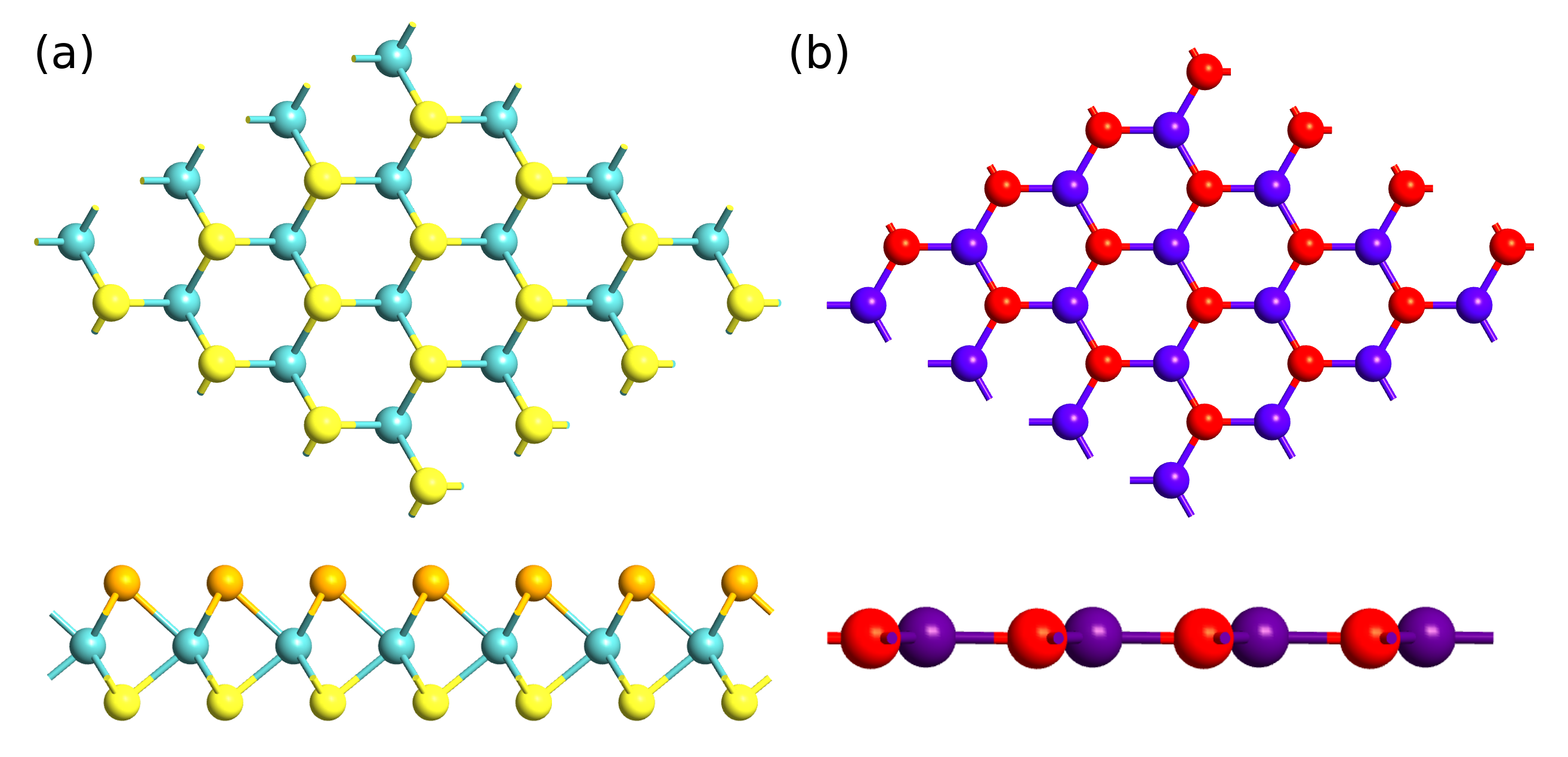}}
\caption{(Colour online) The top and side views of (a) monolayer MoSSe  and (b) monolayer AlN. The green, yellow, orange, purple, and red balls represent Mo, S, Se, Al, and N atoms, respectively.}
\label{ml_structure}
\end{figure} 
We  first   started  investigating   the  structural   and  electronic properties of  MoSSe and AlN  monolayer. The atomic  configurations of  MoSSe and  g-AlN monolayers are  shown  in figure   \ref{ml_structure}.  They  have a  two dimensional honeycomb-like hexagonal lattice with  the optimized lattice constants $a$ of 3.248 {\AA}  and 3.169 {\AA} for  MoSSe and  g-AlN monolayers, respectively, with a good agreement of experimental and theoretical works \cite{theo-AlN, exp-AlN, mosse}.  Table \ref{tab1} summarizes a comparison of experimental and theoretical structural parameters of monolayered MoSSe, g-AlN, and SMoSe/g-AlN heterostructure.

\begin{figure}[!t]
	\centering{\includegraphics[width=14cm]{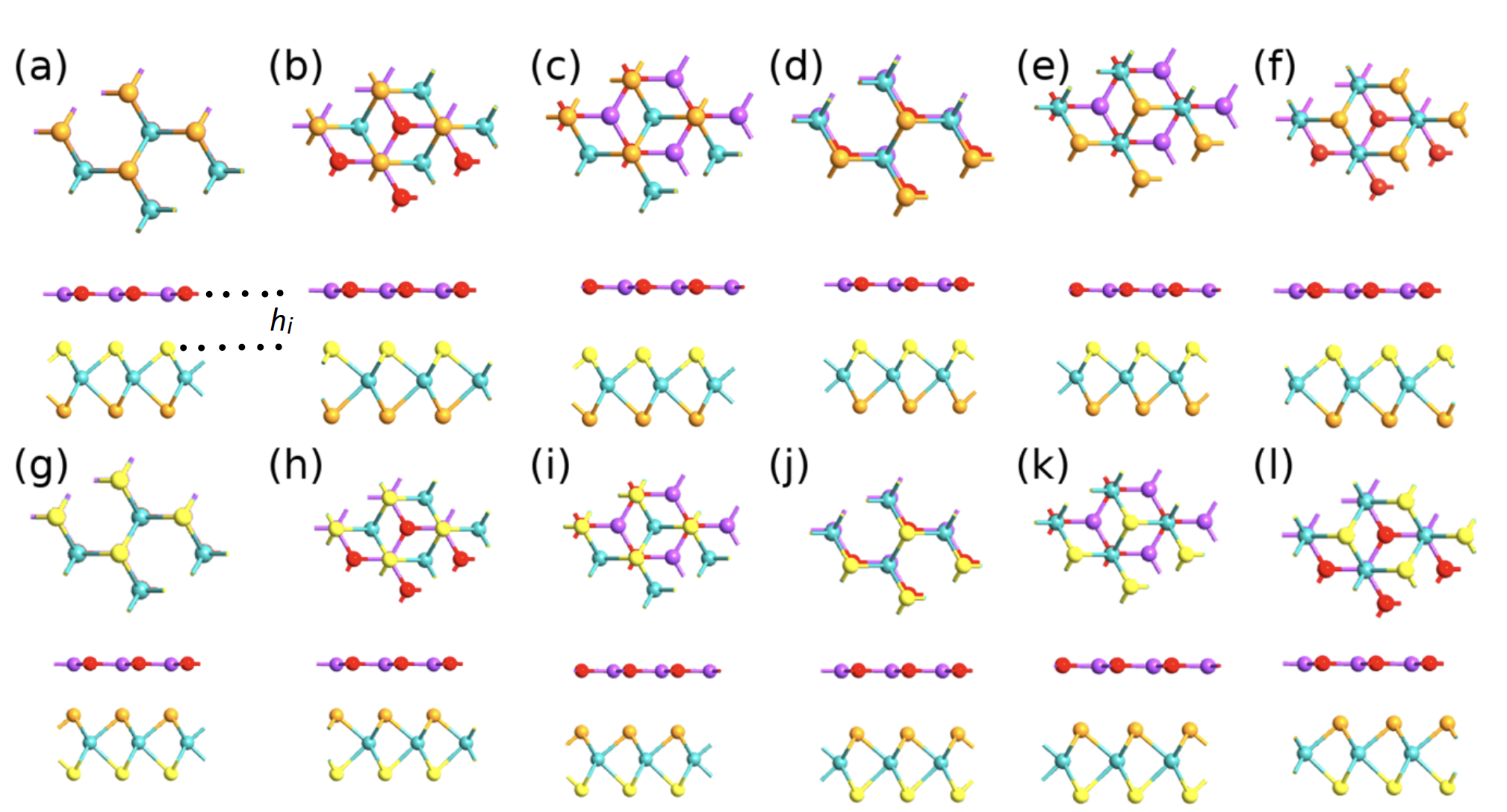}}
	\caption{(Colour online) The atomic structure of the different stacking (a-l) heterostructures in the type of SeMoS/g-AlN and SMoSe/g-AlN. The top and side views are in the upper and lower panels, respectively. The green, yellow, orange, purple, and red balls represent Mo, S, Se, Al, and N atoms, respectively. The interlayer distance between the top and bottom layer is denoted by the letter $h{_i}$.}
	\label{heterostructs}
\end{figure}

Comparing the lattice constant of monolayer g-AlN with monolayer, there is a small  ratio of  lattice mismatch  at  about 2.43$\%$.   It makes  possible candidates  to  form  vertical  van der  Waals  (vdW)  heterostructure without rather high strain which indicates  the capability to synthesize MoSSe/g-AlN heterostructure in actual production.
\begin{table}[!h]
\caption{The calculated lattice constant ($a$), experimental lattice constant ($a_{\rm expt}$) \cite{theo-AlN, exp-AlN, mosse}, bond length ($d$), interface distance ($h{_i}$), formation energy ($E_f$), and bandgap of the monolayered MoSSe, g-AlN, and SMoSe/g-AlN heterostructure. The band-gap values are labeled by letter D or I, which represents the direct or indirect band gap.}
\vspace{3mm}
\label{tab1}
\centering{\begin{tabular} {p{4.5cm} p{3cm} p{3cm} p{3cm}}
 \hline  \hline
 Property/System &MoSSe& g-AlN & SMoSe/g-AlN \\ \hline \hline
$a$~({\AA}) & 3.248 &3.169  & 3.169 \\ 
$a_{\rm expt}$ &3.160 \cite{mosse} &3.110 \cite{theo-AlN, exp-AlN} & -- \\ 
$d_{{\rm Mo-S}}$~({\AA}) & 2.427&-- &2.404 \\ 
$d_{{\rm Mo-Se}}$~({\AA}) &2.542 & --& 2.520 \\ 
$d_{{\rm Al-N}}$~({\AA}) & --&1.829 & 1.829 \\ 
$h_{i}$~({\AA}) &-- &-- &3.013 \\ 
$E_f~({\rm meV/atom}) $ &-- &-- & 99.282 \\ 
$E_g~({\rm eV})$ & 1.502 (D)&3.455 (I) &1.627 (I) \\ \hline \hline
\end{tabular}}
\end{table}
Due to the different faces of monolayer MoSSe, we considered 12 different configurations to find  the most  stable  structure. For  SMoSe/g-AlN heterostructure,  we considered the following configurations as shown in figure  \ref{heterostructs}(a-f): (a) Mo atom on top of the N atom, and S  atom on top of  the Al atom (Mo--N);  (b) Mo atom on  top of the hole site and S atom on top of the  Al atom (Mo--H Al); (c)~Mo atom on top of the hole site and S atom on  top of the N atom (Mo--H N); (d) S atom on top of the N atom and Mo atom on top of the Al atom (S--N); (e) S atom on top of the hole site and Mo atom on top of the N atom (S--H N); (f) S  atom on top of  the hole site and  Mo atom on top  of the Al atom  (S--H Al). The  similar configurations  are also  used for  the SMoSe/AlN heterostructure with Se, not  S, atoms facing the underlying g-AlN layer  as shown in figure  \ref{heterostructs}(g-l).  From the previous studies, it is found that the MoSSe/AlN heterostructures showed the lowest 	interaction energies when the MoS face of MoSSe was interacting with the AlN sheet. 	This suggests that the sulfur component in MoSSe exhibits a higher degree of reactivity 	and contributes greatly to the interaction with AlN \cite{chem}.
To evaluate the  relative stability  of 12  representative highly  symmetrical
geometrical structures, we calculated the formation energy, $E_{\rm f}$, which is given by the following formula,
\begin{equation}
	E_{\rm f}=E_{\rm MoSSe/g-AlN}-E_{\rm MoSSe}-E_{\rm g-AlN},
\end{equation}
where the $E_{\rm MoSSe/AlN}$, $E_{\rm MoSSe}$, and $E_{\rm g- AlN}$ are  the  total  energies  of  the  MoSSe/g-AlN heterostructure, MoSSe  and g- AlN  monolayers, respectively.  From our results, we found that SMoSe/g-AlN stacking (Mo--N  configuration) has the lowest $E_{\rm f}$, of 99.282 meV/atom, suggesting this arrangement to  be  energetically   competitive  and potentially accessible by experiments. The calculated interlayer distance, $h_{i}$ is 3.013~{\AA}  for this configuration. Both the interlayer distance (3.336~{\AA} \cite{grap}) and the formation energy (18~mev/{\AA}$^{-2}$~\cite{che}) are smaller than those in the vdW bonding in weak interlayer interactions in graphite. This indicates that the SMoSe/g-AlN heterostructues were formed by the vdW force that exists between the layers. Furthermore, the interlayer distance demonstrates a similarity to both  bilayer MoS$_2$ and MoSe$_2$ which are 3.12 {\AA} and 3.13~{\AA}, respectively \cite{ms,mse}.
\begin{figure}[!t]
	\centering{\includegraphics[width=13cm]{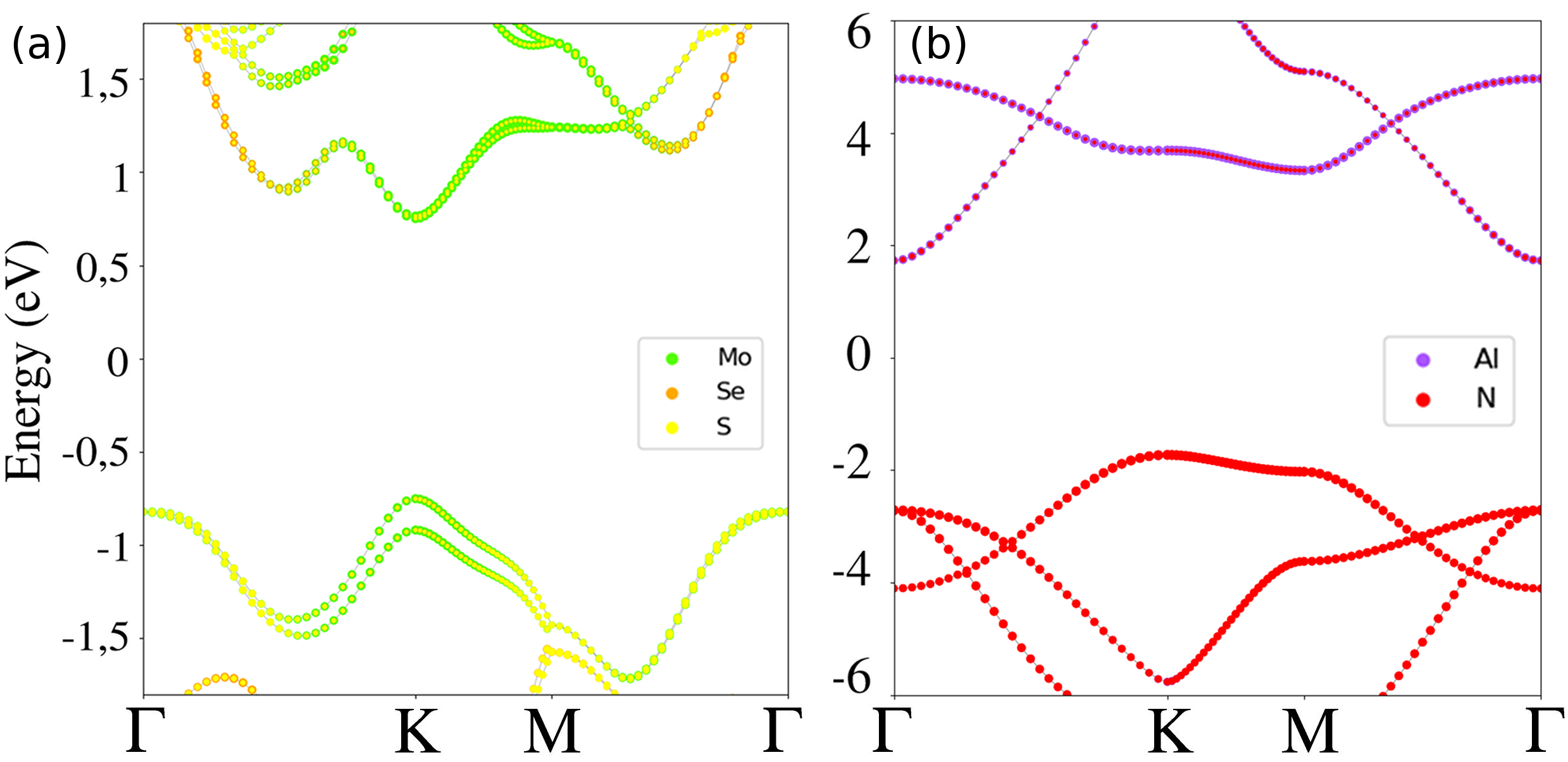}}
	\caption{(Colour online) The projected band structures of (a) the prefect monolayer MoSSe and (b) monolayer g-AlN obtained at PBE levels including spin-orbit coupling. Fermi level is set to zero energy.}
	\label{bs_ml}
\end{figure} 
\begin{figure}[!t]
\centering{\includegraphics[width=13cm]{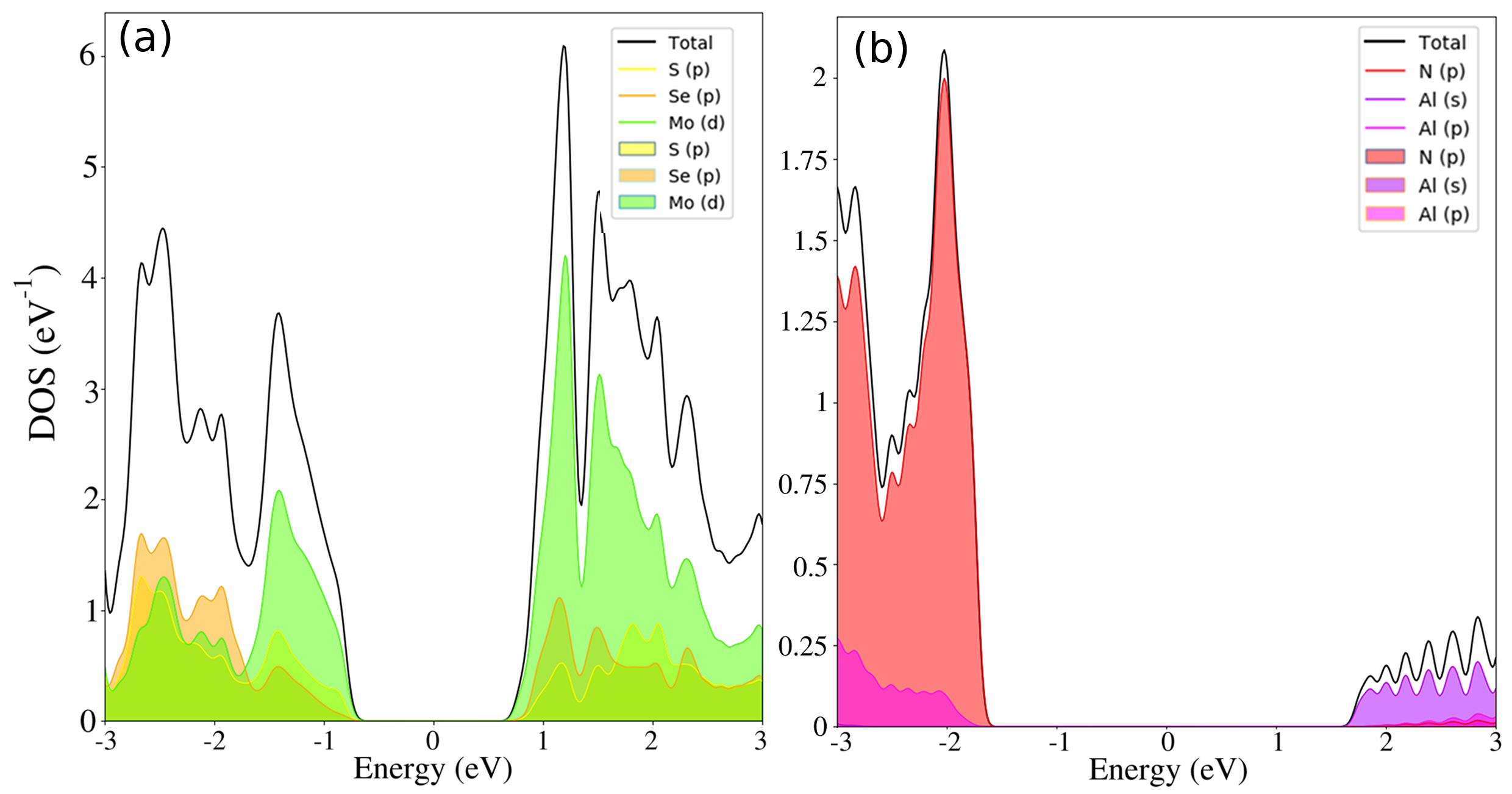}}
	\caption{(Colour online) The partial density of states for the (a) Janus MoSSe monolayer and (b) g-AlN monolayer. The total density, partial density of Mo $d$, partial density of S $p$, partial density of Se $p$, partial density of Al $s$, partial density of Al  $p$, and partial density of N $p$ are in black, green, yellow, orange, light purple, purple and red colors, respectively. Fermi level is set to zero energy.}
	\label{dos_ml}
\end{figure}

\begin{figure}[!t]
	\centering{\includegraphics[width=14cm]{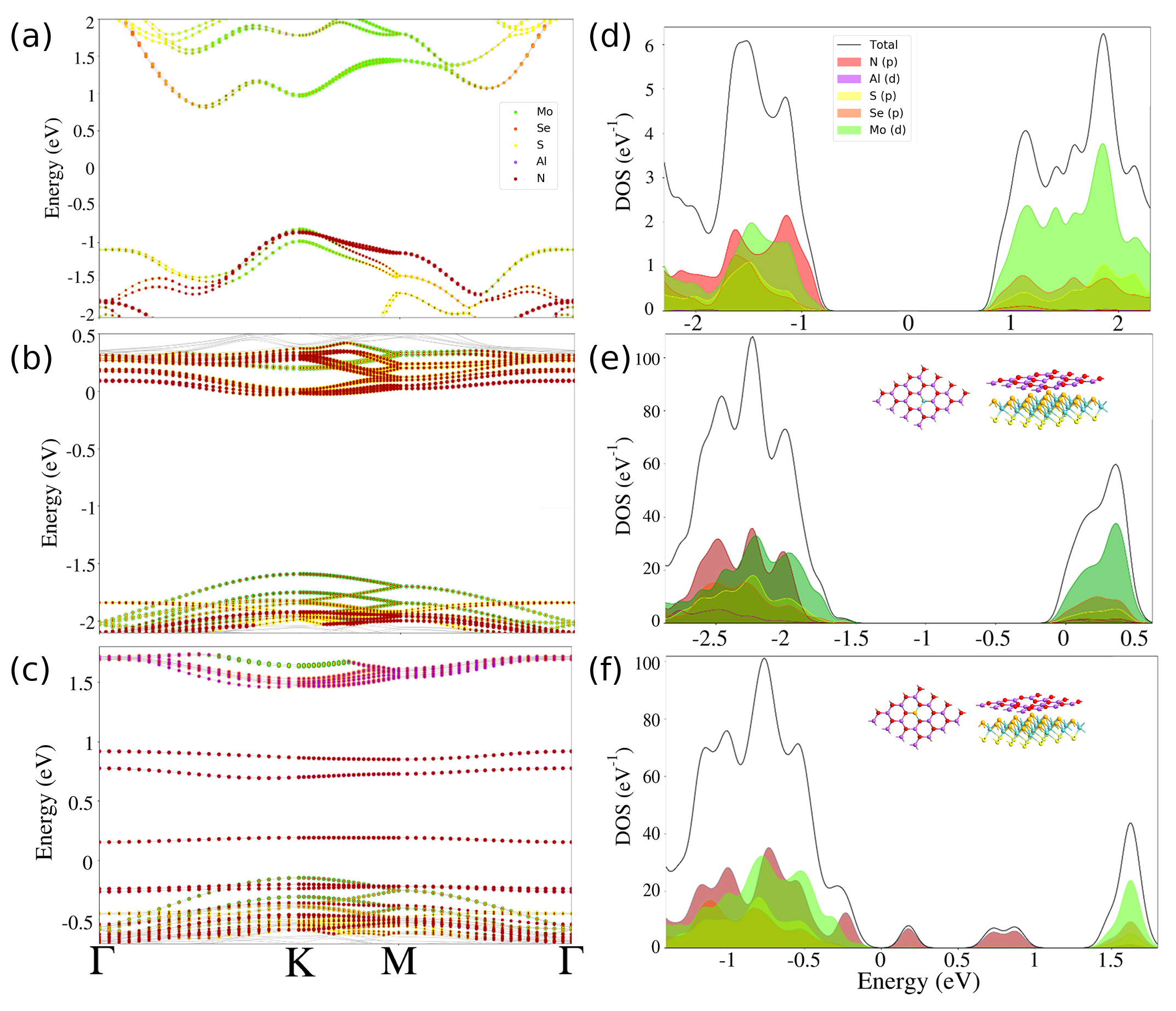}}
	\caption{(Colour online) The projected band structures of (a) MoSSe/g-AlN, (b)  MoSSe/g-AlN-V$_{\rm N}$, and (c) MoSSe/g-AlN-V$_{\rm Al}$ obtained at PBE levels including spin-orbit coupling. The partial density of states for the (d) SMoSe/h-AlN, (e)  SMoSe/g-AlN-V$_{\rm N}$, and (f) MoSSe/g-AlN-V$_{\rm Al}$ heterostructures. The optimized structures of SMoSe/g-AlN-V$_{\rm N}$ and  MoSSe/g-AlN-V$_{\rm Al}$  are shown as an inset, respectively. The total density, partial density of Mo $d$, partial density of S $p$, partial density of Se $p$, partial density of Al  $p$, and partial density of N $p$ are in black, green, yellow, orange, purple and red colors, respectively. Fermi level is set to zero energy. }
	\label{bs_hetero_vn}
\end{figure} 

Figure \ref{bs_ml}(a) and (b) displays projected band structures of monolayer MoSSe and g-AlN structures. It is shown that the both structures are all semiconductors. For the MoSSe monolayer, both the valence band maximum (VBM) and the conduction band minimum (CBM) are located at the {\bf K} point, suggesting that  this structure is a direct semiconductor with a band gap value of 1.502 eV. The both VBM and CBM are mostly contributed by Mo atoms as shown in figure  \ref{bs_ml}(a). However, the monolayer g-AlN is an indirect semiconductor with the band gap value of 3.455 eV in which the VBM and CBM are located at {\bf K} and {\bf $\Gamma$} points, respectively, which have a good agreement with previous studies \cite{guan, ren, yang}. As presented in figure  \ref{bs_ml}(b), the VBM is contributed by N atoms whereas the CBM is contributed by Al atoms. We also calculated the partial density of states (PDOS) of both monolayer MoSSe and g-AlN structures to investigate the atomic contribution to VBM and CBM, as displayed in figure  \ref{dos_ml}(a) and (b), respectively. For the monolayer MoSSe, the  VBM and CBM are mainly contributed by Mo $d$ orbitals, whereas the VBM of monolayer g-AlN is mainly contributed by N $p$ orbitals and the main contribution to the CBM stem from the Al $s$ orbitals.

Next, we use the most stable SMoSe/g-AlN heterostructure to follow our investigations. Figure~\ref{bs_hetero_vn}(a) shows the electronic structure of the vertical SMoSe/g-AlN heterostructure. It is found that the SMoSe/AlN heterostructure is an indirect semiconductor with the band gap of 1.627 eV which is in the visible-infrared region. The VBM is mainly contributed by Mo and N atoms. Whereas there are less contributions from N and S atoms, the most contributions come from Mo and Se atoms for CBM. We also report the calculated PDOS to clarify the atomic contributions. As shown in figure~\ref{bs_hetero_vn}(d), the main contribution to VBM arises from N $p$ orbitals. On the contrary, the main contributions to the CBM are due to the Mo $d$  and Se $p$ orbitals of MoSSe monolayer.

To explore different defect states in the most stable SMoSe/g-AlN heterostrcutre, we built the 4$\times$4 supercell SMoSe/g-AlN heterostructure to form a vacancy by removing Al or N atom (V$_{\rm Al}$ and V$_{\rm N}$) in monolayer g-AlN.  We introduced Al and N single vacany defects in the (4$\times$4) supercell of g-AlN monolayer. The Al and N atoms are removed from the center of g-AlN monolayer as shown in the inset of figure~\ref{bs_hetero_vn}(e--f). The electronic structure of SMoSe/g-AlN-V$_{\rm N}$ and  SMoSe/g-AlN-V$_{\rm Al}$ is shown in figure~\ref{bs_hetero_vn}(b) and (c), respectively. One can see that several defect states occur within the SMoSe/g-AlN heterostructure by introducing  an Al vacancy. However, when the N vacancy is created in SMoSe/g-AlN heterostructure, the semiconductor nature is preserved with  the Fermi level slighlty shifted to CBM making n-type material. As displayed in \ref{bs_hetero_vn}(b), the VBM is mainly determined by Mo atoms whereas the CBM is mainly determined by N atoms for the SMoSe/h-AlN-V$_{\rm N}$ heterostructure. Conversely, with the Al vacancy, there are many defect states in the band gap which mainly originate from N atoms. To better investigate this, the partial density of states is shown in figure  \ref{bs_hetero_vn}(d)--(f). For the defect free SMoSe/g-AlN heterostructure,the VBM is derived from N $p$ orbital and the CBM is derived from Mo $d$ orbital. When the N vacancy is introduced, the VBM is mainly determined by both Mo $d$ and N $p$ orbitals, although the CBM is still determined by Mo $d$ orbital. It is worth noting that for  all heterostructures the CBM consists of Mo-$d$ orbitals. Moreover, the only active defect states in SMoSe/g-AlN-V$_{\rm Al}$  mainly originate from N $p$ orbital. They are located close to the valance band. They are also optically active defects which act as an acceptor. These active defects are loacated in the band gap region of~1.358 eV which is in the visible-infrared region, indicating that this heterostructure is suitable for nanoscale sensing application.

The van der Waals interface can be perceived as a two-dimensional defect and the overall material properties are strictly related to those of each component \cite{shab}. Defects at interfaces can significantly impact the the original electrical, optical, vibrational, magnetic, and chemical properties of materials. They can introduce localized states within the band gap, affecting the band alignment and charge transfer across the interface. These defects are inevitably introduced during the fabrication of material. They are particularly relevant in heterostructures used for electronic and optoelectronic applications such as vertical rectifying junctions, photodiodes  and resonant tunneling diodes \cite{leec, liny} . In certain cases, defects provide benefits to material properties, enhancing the device performance, and enabling unprecedented functionalities. To understand and control the defects at interfaces is essential for optimizing the device performance. The charge transition levels of defects in TMDs were studied experimentally \cite{barja,wu} and computationally \cite{koms, tan}. Controlling the formation of the defects in the SMoSe/g-AlN heterostructure can lead to new electronic properties in the heterostructure  creating opportunities to expand their potential applications. Furthermore, our results show that the Al or N vacancy in this structure can offer a route for manipulating the band gap in spintronics and optoelectronics, allowing for new possibilities in these fields. We also found that the Fermi level can be tuned by introducing a N vacancy  in the heterostructure. The findings demonstrate that the presence of Al or N vacancy in g-AlN enhances the interaction of the heterostructure. Our theoretical calculations play a significant role in accurately assessing the effect of defects on material properties, which helps to explain experimentally the observed phenomena.

\section{Conclusions}
In conclusion, we examined the structural and electrical characteristics of SMoSe/g-AlN heterostructures with and without vacancies. The SMoSe/g-AlN structure maintains a flat configuration for its both sublayers in the absence of defects. The introduction of Al and N vacancies, on the other hand, causes localized distortions. The results demonstrate that SMoSe/g-AlN (Mo-N configuration) is more stable than other forms of stacking, with a binding energy of 99.282 meV/atom, independent of possible atomic locations. The stable SMoSe/g-AlN heterostructure has an indirect band gap of 1.627 eV based on the results of the projected band structure, which is in the visible infrared range. The heterostructure exhibits comparable electrical band topologies with localized states inside the band gap by creating vacancies in the substrate g-AlN. Furthermore, due to the asymmetry of the MoSSe monolayer, this heterostructure has an inherent dipole moment perpendicular to the two-dimensional plane, which may be exploited to tune the interfacial interactions. The research findings obtained might give useful insights for utilization in nano/optoelectronic devices.

\section{Acknowledgement}

\"O. C. Yelgel acknowledges the support from the University of Manchester, National Graphene Institute and School of Physics \& Astronomy.

\newpage
\ukrainianpart

\title{Роль власних атомних дефектів у гетероструктурі Януса MoSSe/XN (X = Al, Ga): дослідження за першими принципами}
\author{О. С. Єлгел} 

\address{
	Університет Реджепа Тайїпа Ердогана, факультет електротехніки та електроніки, 53100, Різе, Туреччина}

\makeukrtitle

\begin{abstract}
	Взаємодії між різними шарами в гетероструктурах Ван-дер-Ваальса мають значний вплив на електронні та оптичні характеристики. Ці міжшарові взаємодіє можна налаштовувати, використовуючи власний дипольний момент дихалькогенідів перехідних металів Януса (TMD). Автори систематично досліджують структурні та електронні властивості гетероструктур Януса MoSSe ``моношар/графеноподібний нітрид алюмінію (MoSSe/g-AlN) з точковими дефектами'', використовуючи розрахунки теорії функціоналу густини з включенням нелокальної поправки Ван-дер-Ваальса. Отримані дані свідчать про те, що досліджувані гетероструктури є енергетично та термодинамічно стійкими, а їх електронні структури можна легко модифікувати шляхом створення гетероструктури з дефектами моношару g-AlN. Ця гетероструктура є непрямим напівпровідником із забороненою зоною 1.627~еВ, яка знаходиться у видимій інфрачервоній області. Це може представляти інтерес для створення фотоелектричних пристроїв. Коли один атом N або Al видаляється з моношару g-AlN у гетероструктурі, створюючи дефекти вакансії, матеріал демонструє подібні електронні зонні структури з локалізованими станами в межах забороненої зони, які можна використовувати для планомірного налаштування електронних властивостей гетероструктури MoSSe/g-AlN. 
	Ці налаштування можуть запропонувати широкі можливості для розробки наноелектронних пристроїв, що базуються на  гетеропереходах MoSSe/g-AlN.
	\keywords гетероструктури, квантовий еспресо, структури Януса, теорія функціоналу густини
\end{abstract}

\end{document}